\documentclass{PoS}

\title{Color singlet and adjoint free energy\\at finite temperature }

\ShortTitle{Color singlet and adjoint free energy at finite temperature }

\author{\speaker{Alexei Bazavov}$^a$, P\'eter Petreczky$^b$,
        Alexander Velytsky$^{c,d}$\\
\llap{$^a$} Department of Physics, University of Arizona, Tucson, AZ 85721, USA\\
\llap{$^b$} RIKEN-BNL Research Center and Physics Department, Brookhaven
  National Laboratory, Upton NY 11973, USA\\
\llap{$^c$} Enrico Fermi Institute, University of Chicago, 5640 S. Ellis Ave., Chicago, IL 60637, USA\\
\llap{$^d$} HEP Division and Physics Division, Argonne National Laboratory, 9700 Cass Ave., Argonne, IL 60439, USA
}

\abstract{
We study correlation functions of static quark-antiquark pairs 
in $SU(2)$ gauge theory at finite temperature. By measuring 
Polyakov loop correlators and temporal Wilson loops with 
APE smearing of spatial links we are able to give consistent 
definitions of the singlet and adjoint free energies at short
distances, where the notion of these free energies is meaningful. 
APE smearing procedure allows to achieve a high degree of overlap 
in the singlet channel and to reconstruct the adjoint part from the
color averaged and singlet free energy.
}

\FullConference{The XXVI International Symposium on Lattice Field Theory \\
                 July 14 - 19, 2008\\
                 Williamsburg, Virginia, USA}

\begin{document}

\section{Introduction}
Strongly interacting matter undergoes a 
deconfining transition at some temperature which is triggered by large
increase in the number of degrees of freedom 
as well as melting of hadronic states.
Another important feature of the deconfined phase is the screening
of color charges. 
It has been argued that color screening will lead to quarkonium dissociation
above the deconfinement temperature which can be used as a signature of
quark gluon plasma formation in heavy ion collisions \cite{MS86}.
Melting of quarkonium states has been studied in potential models with
screened potentials (see e.g. \cite{mocsy} and references therein).
Alternatively, one may try to reconstruct quarkonium spectral functions from
Euclidean correlators (see e.g. \cite{jako,datta04} and references therein).
It turns out, however, that quarkonium melting does not affect the Euclidean meson
correlators and these correlators are almost temperature independent. 
This is contrary to what happens in the light quark sector, where both meson correlators
and spectral functions show significant temperature dependence in the deconfined phase \cite{karsch02}.
Therefore potential models are useful tools to study in-medium quarkonium properties. But to establish
their applicability a better understanding of color screening is needed.

On the lattice color screening is usually studied in terms of the Polyakov
loop correlator related to the free energy of static quark anti-quark pair \cite{mclerran81}.
This correlator shows significant temperature dependence across the
transition and in the deconfined phase the free energy of static 
quark anti-quark pair shows
large temperature dependence even for very small separations between the static quark and anti-quark,
much smaller than the inverse temperature. 
In perturbative picture this can be understood due to the fact that in the deconfined phase not only
singlet quark anti-quark ($Q\bar Q$) states contribute to the free energy but also colored states with
$Q\bar Q$ in the adjoint representation. 
This observation is also supported by lattice
calculations of the correlation function of two temporal Wilson lines in Coulomb gauge, which in 
perturbation theory corresponds to the so-called singlet free energy \cite{mehard}.
 The singlet free energy is temperature
independent at short distances and coincides with the zero temperature potential as expected. At
large distances, however, it also shows significant temperature dependence.

The problem of defining color singlet and adjoint $Q\bar Q$ states
on the lattice has been considered in Ref. \cite{jahn}. 
It has been found that the conventional definition of singlet and
adjoint states have problems. 
Here we report on our study \cite{BPV2008} of
static meson correlators in 4 dimensional $SU(2)$ gauge theory
at finite temperature and show how the problem observed in 
Ref. \cite{jahn} can be resolved in the limit of small distances and/or
high temperatures.

\section{Static meson correlators}

On the lattice 
correlators of the static meson operators 
in color singlet and adjoint states at $t=1/T$
have the form \cite{jahn}:
\begin{eqnarray}
&
\displaystyle
G_1(r,T)=\frac{1}{N} 
\langle {\rm Tr} 
L^{\dagger}(x) U(x,y;0) L(y) U^{\dagger}(x,y,1/T) \rangle, 
\,\,\,\,\,r=|x-y|,\label{defg1}\\[2mm]
&
\displaystyle
G_a(r,T)=\frac{1}{N^2-1}
\langle {\rm Tr} L^{\dagger}(x)  {\rm Tr} L(y) \rangle
-\frac{1}{N (N^2-1)} 
\langle {\rm Tr} L^{\dagger}(x) U(x,y;0) L(y) U^{\dagger}(x,y,1/T) \rangle, \label{defg3}\end{eqnarray}
where $N$ is the number of colors ($N=2$ in our
numerical calculations).
Here $L(x)$ is the temporal Wilson line, which on the lattice is simply $L=\prod_{\tau=0}^{N_\tau-1} U_0(x,\tau)$
with $U_0(x,\tau)$ being the temporal links.
The correlators depend on the choice of the spatial transporters
$U(x,y;t)$. 
In the special gauge, where $U(x,y;t)=1$ the above
correlators give standard definition of the singlet and
adjoint free energies  
\begin{equation}
e^{-F_1(r,T)/T}=
\frac{1}{N} \langle {\rm Tr} L^{\dagger}(x)  L(y)\rangle,\,\,\,\,\,\,\,\,
e^{-F_a(r,T)/T}=
\frac{\langle {\rm Tr} L^{\dagger}(x)  {\rm Tr} L(y) \rangle}{N^2-1}
-\frac{\langle {\rm Tr} L^{\dagger}(x)  L(y)\rangle}{N (N^2-1)}.
\end{equation}
The singlet and triplet free energies can be calculated at high temperature in
leading order HTL approximation \cite{mehard}.
At this order $F_1$ and $F_a$ are gauge independent or in other words do not depend
on the choice of
the parallel transporters $U(x,y;t)$.
At small distances 
the singlet free energy is temperature independent 
and coincides with the zero temperature potential, while the adjoint free energy 
depends on the temperature \cite{mehard}.

The physical free energy of a static 
quark anti-quark pair is given by the thermal average
of the singlet and adjoint free energies and
is explicitly gauge independent
\begin{equation}
e^{-F(r,T)/T}=
\frac{1}{N^2} e^{-F_1(r,T)/T} + \frac{N^2-1}{N^2} e^{-F_a(r,T)/T}
=\frac{1}{N^2} \langle {\rm Tr} L(x) {\rm Tr} L(y) \rangle 
\equiv \frac{1}{N^2} G(r,T) \label{defg}.
\end{equation}

Using
the transfer matrix one can show that in the confined phase \cite{jahn}
\begin{equation}\label{g1}
G_1(r,T)=\sum_{n=1}^{\infty} c_n(r) e^{-E_n(r,T)/T},~~~~~~~~
G(r,T) = \sum_{n=1}^{\infty} e^{-E_n(r,T)/T},\label{g}
\end{equation}
where $E_n$ are the energy levels of static quark and anti-quark pair. The 
coefficients $c_n(r)$ depend on the choice of 
the gauge transporters.
The color averaged correlator $G(r,T)$ 
does not contain $c_n$. The lowest energy level is the usual static quark anti-quark
potential, while the higher energy levels correspond to hybrid potentials 
\cite{michael92}. 
If $c_1=1$ the dominant contribution to
$G_a$ would be the first excited state $E_2$, i.e. the lowest hybrid potential which
at short distances is related to the adjoint potential. In this sense $G_a$ is related 
to static mesons with $Q\bar Q$ in adjoint state. 
Numerical calculations show, 
however, that $c_1(r) \ne 1$ and depends on the separation $r$. Thus $G_a$ also
receives contribution from $E_1$ \cite{jahn}. The lattice data seem to suggest that $c_1$ 
approaches unity at short distances in accord with expectations based on 
perturbation theory, where $c_1=1$  up to ${\cal O} (g^6)$ corrections  \cite{brambilla00}. 
Therefore  at short distances, $r \ll 1/T$ the color singlet and color averaged free energy
are related $F(r,T)=F_1(r,T) + T \ln (N^2-1)$.
This relation is indeed confirmed by lattice calculations \cite{okacz02}.

\section{Numerical results}
We have calculated correlation functions of static mesons $G_1(r,T)$ and 
$G(r,T)$ both in  the confined and deconfined phase of $SU(2)$ gauge theory.
The details of the calculations are presented in Ref.~\cite{BPV2008}.
We have studied the color singlet and averaged correlators
given by Eqs. (\ref{defg1}) and (\ref{defg}). 
The spatial links entering the transporter $U(x,y;0)$ were smeared using APE
smearing, which has been applied iteratively. The weight of the staple in the APE
smeared link was $0.12$.
For $\beta=2.5$ we use spatial links with 10 steps of
APE smearing and unsmeared spatial links. For $\beta=2.7$
we used unsmeared spatial links as well as spatial links with 10 steps and 20 steps of APE smearing.

\subsection{Color averaged correlator in the confined phase}

\begin{figure}
\begin{center}
\includegraphics[width=0.47\textwidth]{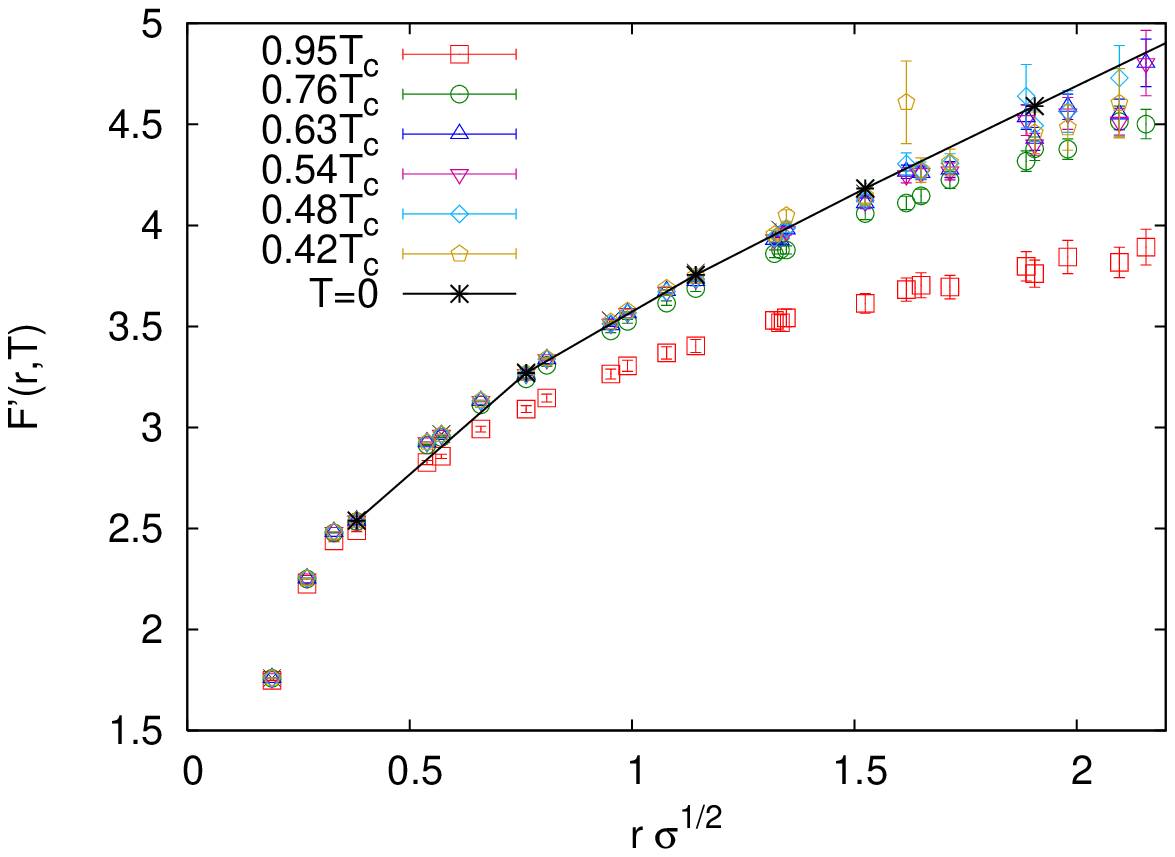}
\hfill
\includegraphics[width=0.47\textwidth]{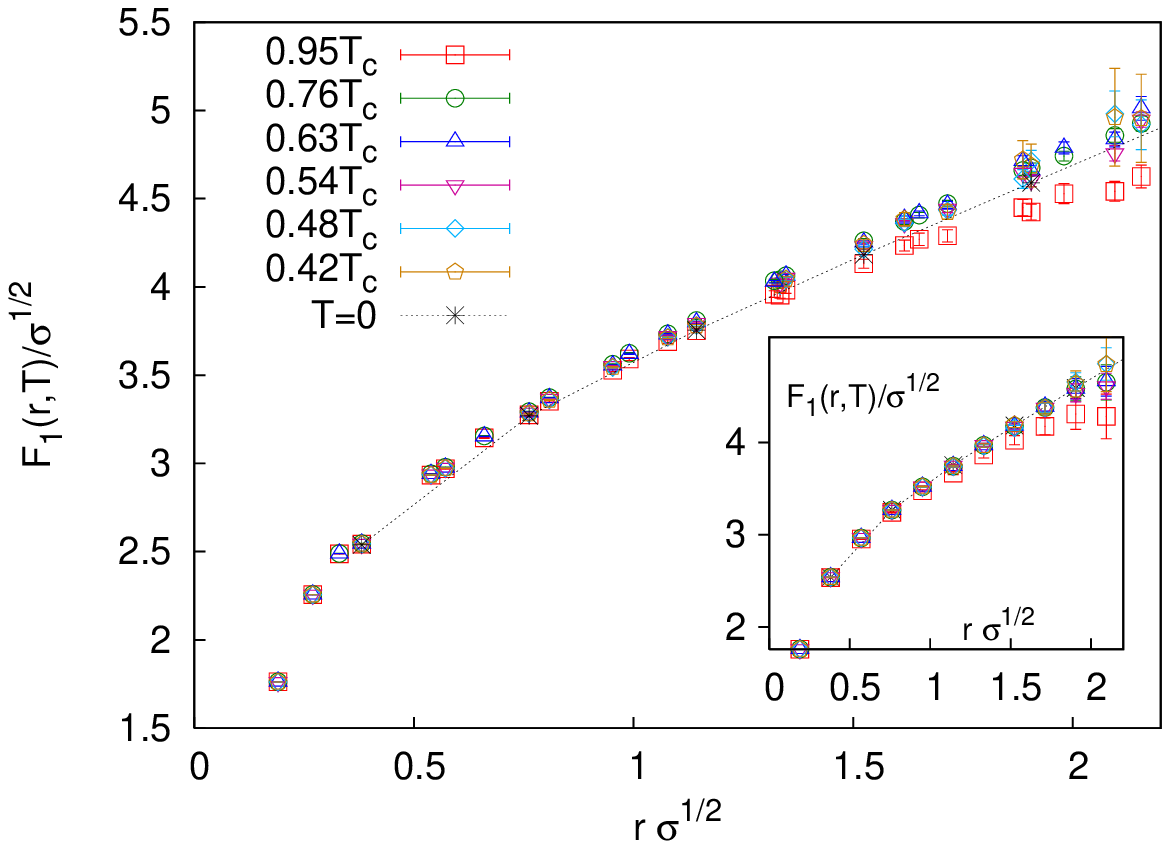}
\parbox[b]{0.47\textwidth}{\caption{The color averaged 
free energy below deconfinement
temperature at $\beta=2.5$ calculated on $32^3 \times N_{\tau}$ lattices.
Also shown is the $T=0$ potential.}\label{fig:fav25}}
\hfill
\parbox[b]{0.47\textwidth}{\caption{The color singlet free energy 
below deconfinement
temperature at $\beta=2.5$ calculated on $32^3 \times N_{\tau}$ lattices.
It is also shown with the contribution 
$T \ln c_1$ subtracted (inset).}\label{fig:f125}}
\end{center}
%\vspace{-3mm}
\end{figure}

The color averaged correlator has been calculated in the
confined phase in the temperature interval $0.32T_c-0.95T_c$ for 
$\beta=2.5$ and $0.49T_c-0.98T_c$ for $\beta=2.7$. 
The numerical results for the color averaged free energy for $\beta=2.5$ are
shown in Figure \ref{fig:fav25}. To eliminate the trivial temperature dependence
due to the color trace normalization in Figure  \ref{fig:fav25} we show the subtracted
free energy $F'(r,T)=F(r,T)-T \ln 4$ together with 
zero temperature potential. The color averaged free energy does not show any
temperature dependence up to temperatures of about $0.76T_c$.
Since the temperature dependence for $T<0.76T_c$ is relatively small 
we attempted to fit the color
averaged correlator with 
the 1-exponential form $G(r,T)=c_1^a(r) \exp(-E_1(r)/T)$.
The ground state energy $E_1(r)$ extracted from this fit agrees
well with the zero temperature potential calculated in Ref. \cite{michael92},
while the coefficients $c_1^a(r)$ are close to one as expected. 
%Although the deviations of $c_1^a(r)$
%from unity are small, they appear to be statistically significant. 
%They are likely to be due to the contribution from excited states
%as the gap between the ground state and excited states (hybrid potential) gets smaller
%with increasing separation.

\subsection{Color singlet correlators in the confined phase}\label{sub_singlet_conf}

The color singlet correlators
have been calculated using different levels of APE smearing in the spatial gauge connection.
We have found that when no smearing is used the color singlet
free energy, $-T \ln G_1(r,T)$ shows a small but visible temperature dependence.
In particular $F_1(r,T)$ is larger than the $T=0$ potential for intermediate
distances $0.5<r \sqrt{\sigma} <2$. 
The temperature dependence of the singlet free energy is significantly
reduced when APE smearing is applied.
In Figure \ref{fig:f125} we show the color singlet free energy for $\beta=2.5$
and 10 APE smearings. As one can see from the figure 
the color singlet free energy shows much smaller temperature dependence
as we get closer to the deconfinement temperature. 

\begin{figure}
\includegraphics[width=0.47\textwidth]{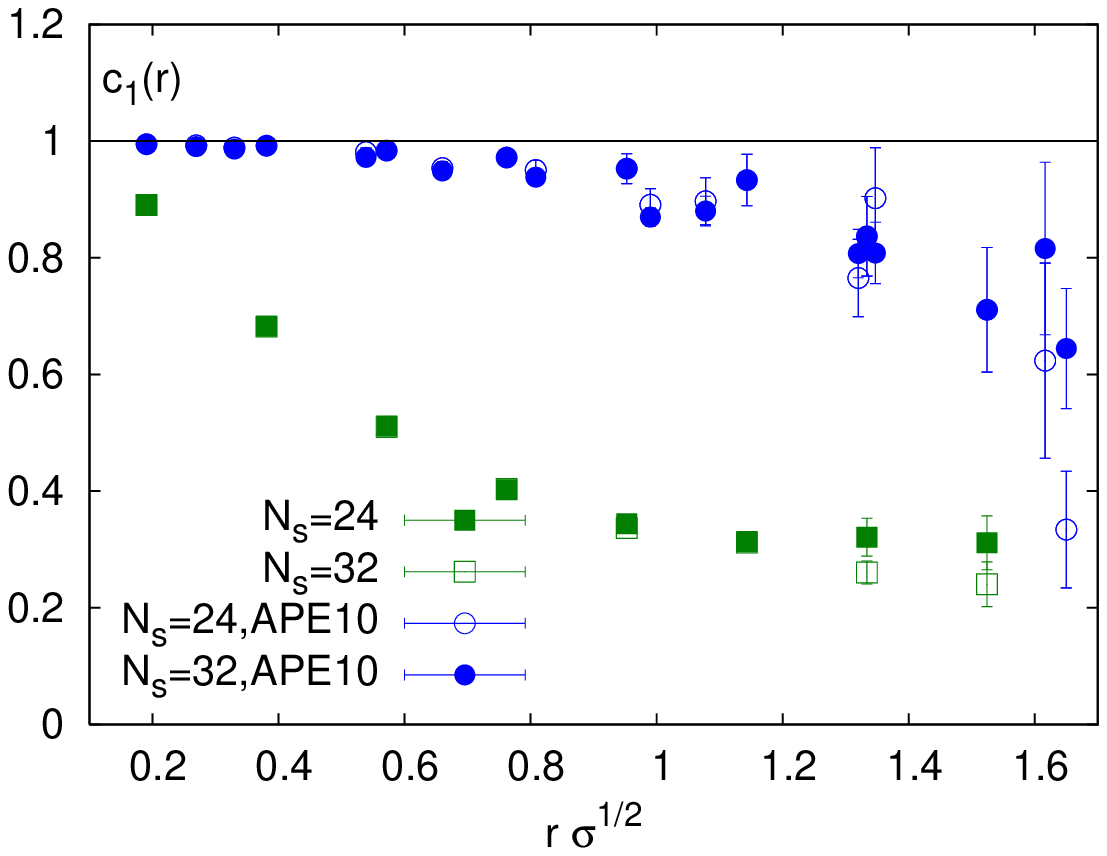}\hfill
\includegraphics[width=0.47\textwidth]{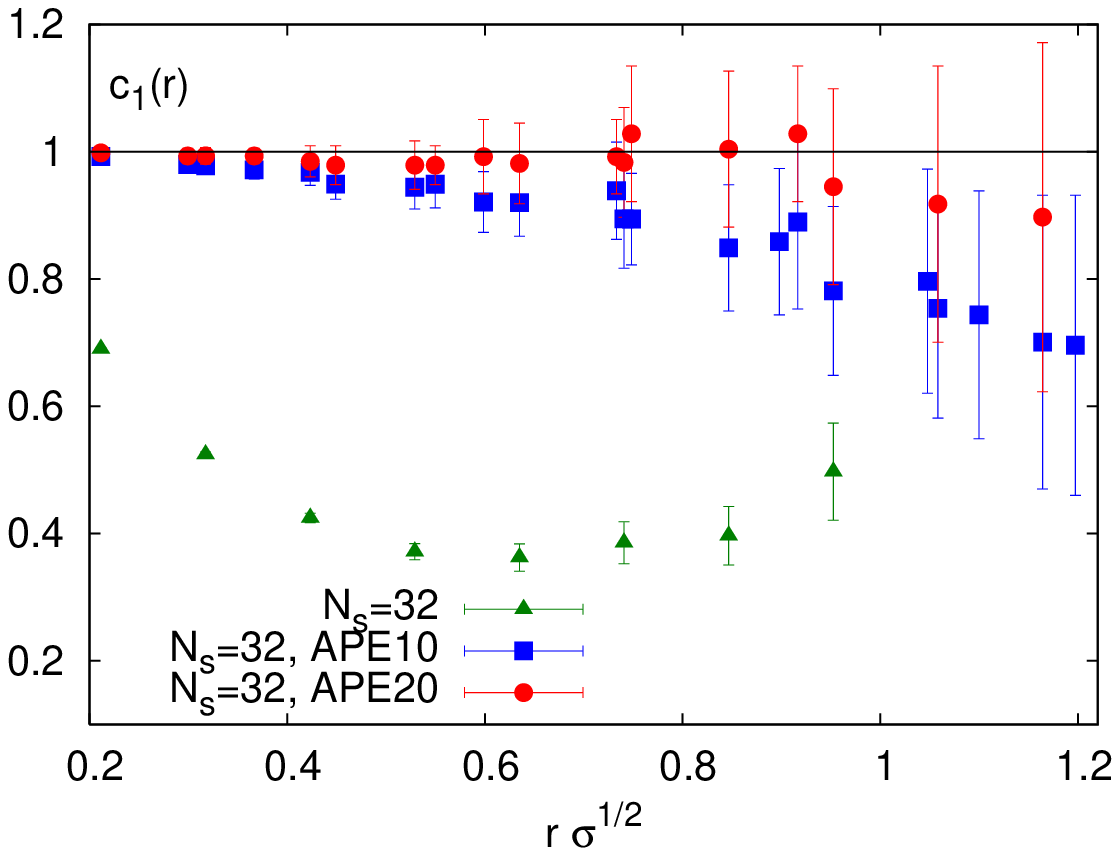}
\caption{The pre-exponential factor of the color singlet correlators as function of
distance $r$ for $\beta=2.5$ (left) and $\beta=2.7$ (right). Shown are results
for unsmeared spatial links and 10 and 20 steps of APE smearing.}
\label{fig:c1}
\end{figure}

To understand the temperature dependence of the color singlet
correlator we use 1-exponential fit $G_1(r,T)=c_1(r) \exp(-E_1(r)/T)$.
In all cases considered the values of $E_1(r)$ extracted from fits are
in good agreement with the calculation of the zero temperature potential in Ref.  \cite{michael92}.
The value of the prefactor $c_1(r)$ is shown in Figure \ref{fig:c1}.
When no APE smearing is used the value of $c_1(r)$ strongly depends on the separation $r$.
At small distances it shows a tendency of approaching unity as one would expect in perturbation theory
and decreases with increasing distance $r$. At large distance its value is around $0.3-0.5$.
Similar results for $c_1$ have been obtained in study of $SU(2)$ gauge theory in 3 dimensions \cite{jahn}.
When APE smearing is applied the $r$-dependence of the amplitude $c_1$ is largely reduced
and its value is close to unity both for $\beta=2.5$ and $\beta=2.7$. For $\beta=2.7$ we also see
that increasing the number of smearing steps from 10 to 20 reduces the deviation of $c_1$ from unity.
In any case at sufficiently short distances $c_1$ is very close to one as expected in perturbation theory.
Thus, almost the entire temperature dependence 
of the singlet free energy at distances $0.5<r \sqrt{\sigma}<2$ is due to the deviation 
of $c_1$ from unity and can be largely reduced by applying APE smearing to
the links in the spatial gauge connections. To further demonstrate this point in the inset of
Figure \ref{fig:f125} we show the results for $F_1(r,T)+T \ln c_1(r)$.

\subsection{Color singlet free energy in the deconfined phase}
\label{sec:deconf_f1}

It turns out that the singlet free energy, $F_1(r,T)=-T \ln G_1(r,T)$,
calculated from cyclic Wilson loops shares the same qualitative 
features as the singlet free energy calculated
in Coulomb gauge \cite{digal03}. At short distances it is
temperature independent and coincides with the zero temperature potential.
At large distances it approaches a constant $F_{\infty}(T)$, 
which  is the free energy of 
two isolated static quarks at infinite separation. 

\begin{figure}
\begin{center}
\includegraphics[width=0.47\textwidth]{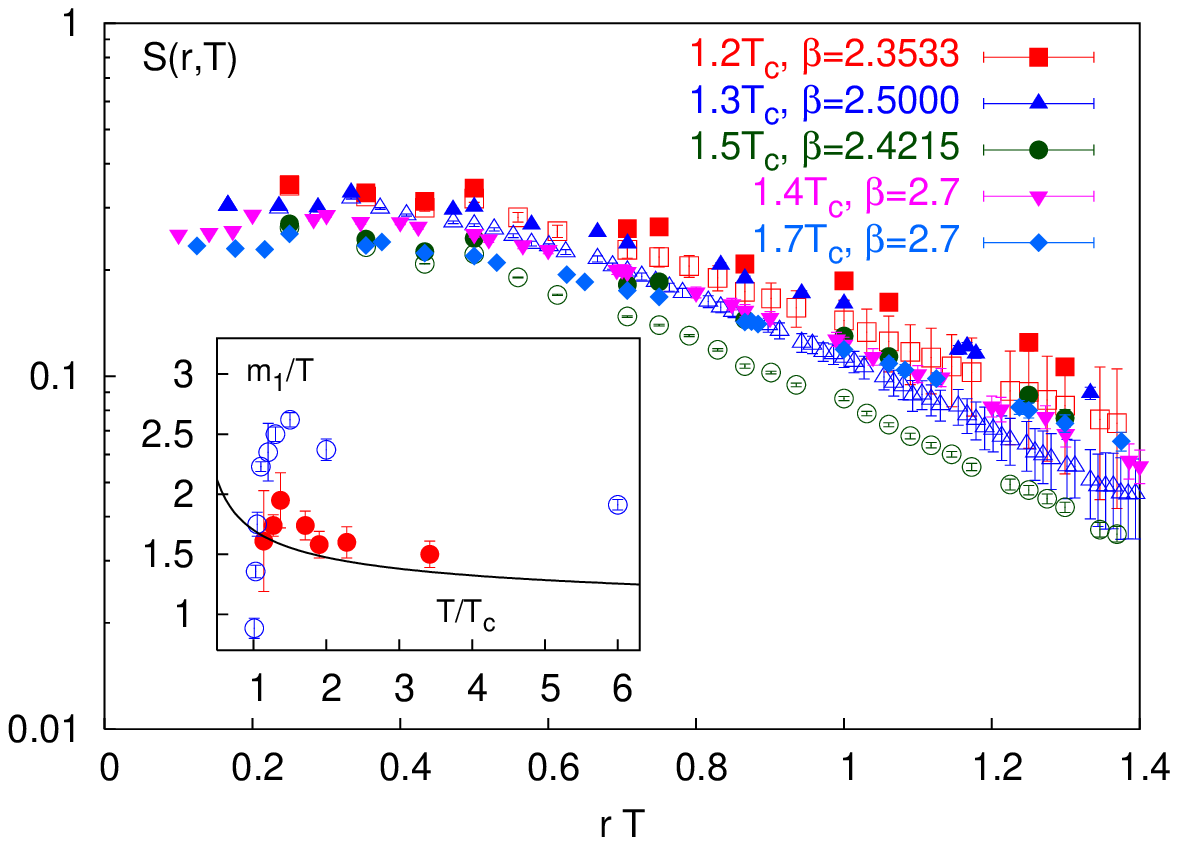}
\hfill
\includegraphics[width=0.47\textwidth]{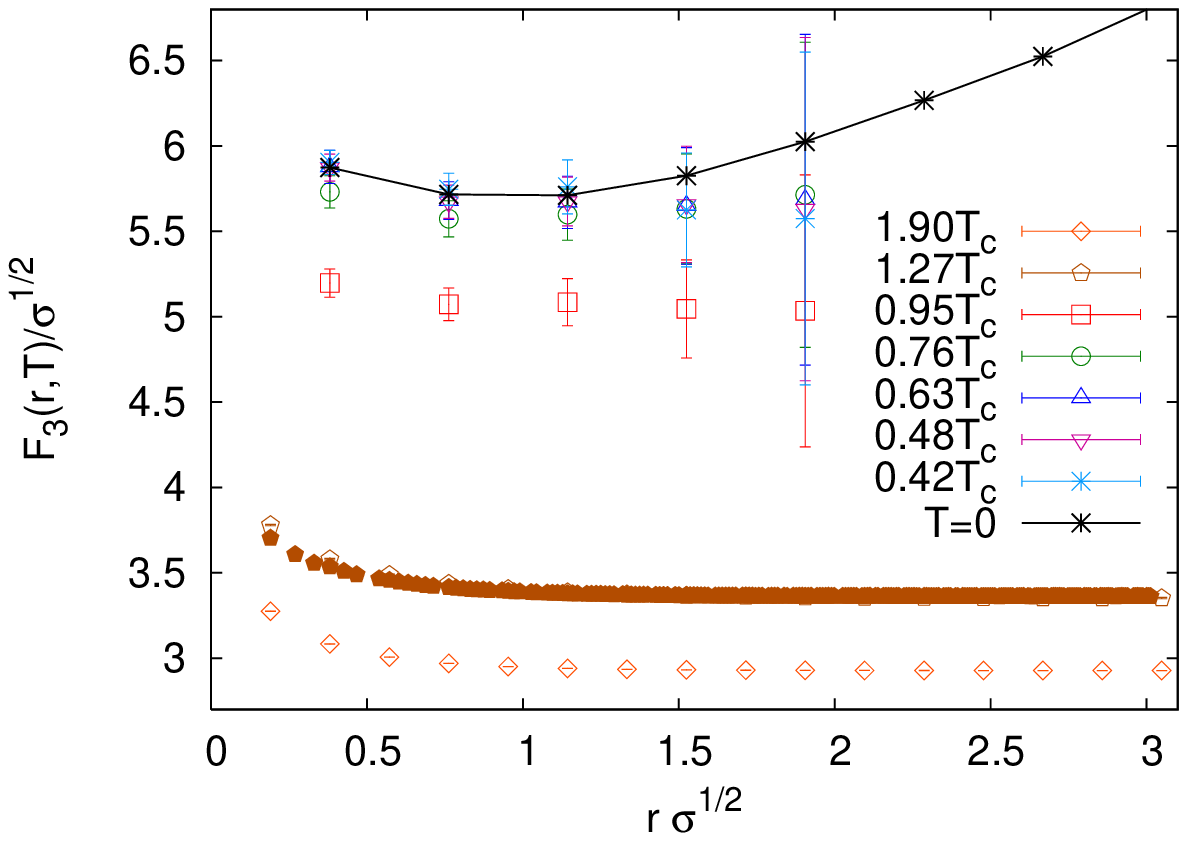}
\parbox[b]{0.47\textwidth}{
\caption{The screening function at different
temperatures as function of $r T$ and different values of $\beta$. 
In the inset the screening masses $m_1$ extracted
from singlet free energies are shown.
The line shows the leading order results for the Debye mass.
} 
\label{fig:s1}
}
\hfill
\parbox[b]{0.47\textwidth}{
\caption{The triplet free energy at different temperatures calculated
at $\beta=2.5$. 
The filled symbols correspond to calculations in Coulomb gauge.
Also shown is the first hybrid potential calculated in Ref. \cite{michael92}.}
\label{fig:f3corr}
}
\end{center}
%\vspace{-3mm}
\end{figure}

At leading order  $F_1(r,T)-F_{\infty}(T)$ is of Yukawa form, therefore
in Figure \ref{fig:s1} we show our numerical results 
in terms of the screening function
$S(r,T)=r \cdot (  F_1(r,T)-F_{\infty}(T))$
at different temperatures.
At short distances ($r T<0.5$) the singlet free energy 
does not depend on the smearing level. Furthermore, it is very close to the free energy
calculated in Coulomb gauge. 
We expect that at large distances the screening function $S(r,T)$ will show 
an exponential decay determined by a temperature dependent screening mass 
$m_1(T)$, which is equal to the leading order Debye mass up to 
the non-perturbative $g^2$ corrections: $m_1=m_D + {\cal O}(g^2)$.
From Fig. \ref{fig:s1} we can see  that indeed $S(r,T)$ behaves exponentially 
with screening mass proportional to the temperature. 
Fitting the large distance behavior of the screening function 
by an exponential form $\exp(-m_1(T) r)$ we
determine the screening mass $m_1(T)$.
In Fig. \ref{fig:s1} we also show the color singlet screening masses 
extracted from the fits and compare them
to the results obtained in Coulomb gauge in Ref. \cite{digal03} as well as to
the leading order Debye mass calculated using 2-loop gauge
coupling $g(\mu=2 \pi T)$ in $\overline{MS}$-scheme. 
As we see from the figure the screening
masses are smaller than those calculated in Coulomb gauge and agree well with the leading order
perturbative prediction.

\subsection{Color triplet free energy}

We have calculated the color triplet correlator defined by Eq. (\ref{defg3})
for different temperatures below and above the transition temperature.
Below the deconfinement temperature we observe a moderate $T$-dependence of the triplet
correlator. We also find that the corresponding free energy $-T\ln G_3(r,T)$ 
is smaller than the first hybrid potential calculated in Ref. \cite{michael92}, but larger than the triplet
free energy  in Coulomb gauge \cite{digal03}. 

Let us assume that only two states contribute to the Eqs. (\ref{g1}).
Then from Eq. (\ref{defg3}) it follows that
\begin{equation}
F_3(r,T)=E_2(r)-T\ln(1-c_2(r)+\frac{1}{3}(1-c_1(r)) e^{\Delta E(r)/T}),
\label{f3_2comp}
\end{equation}
with $\Delta E(r)=E_2(r)-E_1(r)$ and $c_2\ll 1$. 
We have subtracted the correction $T \ln (1+\frac13(1-c_1) e^{\Delta E/T})$ from the triplet
free energy assuming that $E_1(r)$ is given by the ground state potential and $E_2(r)$ is given by the
first hybrid potential as calculated in Ref. \cite{michael92}.
The numerical results are summarized in Fig. \ref{fig:f3corr} which shows that after this 
correction is accounted for in the confined phase the triplet
free energy at low temperatures agrees reasonably well with the first hybrid potential. As temperature
increases more excited states contribute. In particular,  at $0.76T_c$ the value of the triplet
free energy can be accounted for by including the next hybrid state \cite{michael92}.

\section{Conclusions}

We have studied singlet and triplet static quark anti-quark correlators 
in finite temperature $SU(2)$ theory expressed in terms  of Polyakov loop correlators 
and cyclic Wilson loops.
In leading order and probably next-to-leading order of perturbation theory
the static correlators defined by Eq. (\ref{defg1}) and (\ref{defg3}) project
onto singlet and triplet states respectively, however, this separation does not
hold in general case. Due to interactions with ultrasoft fields there will be a mixing
of singlet and triplet states  which is proportional to $\alpha_s^3(1/r)$ and $(r \mu)^4$,
with $\mu$ being the ultrasoft scale \cite{brambilla00}. 
In our case the ultrasoft scale can be the  binding energy, $\alpha_s/r$, $~\Lambda_{QCD}$ or $~g^2 T$.
Therefore it is expected that mixing is quite small at sufficiently small distances.
We determined the mixing between singlet and triplet states in terms of the overlap
factor $c_1(r)$. If the overlap factor is unity there is no mixing. Our lattice
calculations show that $c_1$ indeed approaches one at small distances. 
%Using iterative APE smearing the deviation of $c_1(r)$ from unity can be largely reduced. 
Therefore the contribution of singlet state to $G_3(r,T)$ appears to be small at
temperatures close to deconfinement temperature. This contribution is also
controlled by the non-perturbative gap between the singlet and triplet states, i.e.
the gap between the static potential and the first hybrid potential.

Our analysis shows that at short distances $r T <1 $ the singlet
correlator is almost temperature independent, while the  
triplet correlator is largely affected by the deconfinement.
The temperature dependence of the triplet correlators indicate the melting of the
non-perturbative gap between the singlet and the triplet states above deconfinement,
which turns out to be consistent with perturbative expectations. 
%Because of the
%disappearence of the non-perturbative gap the small deviations of $c_1(r)$ from unity
%play no role in the deconfined phase. In particular, the difference between the
%singlet and triplet correlators defined through cyclic Wilson loops and using
%Coulomb gauge turns out to be quite small. 
This finding is important for application
of thermal pNRQCD discussed in Ref. \cite{brambilla08} to realistic quarkonia and
temperatures not very far from the deconfinement temperature.

\begin{acknowledgments}
This work was supported by U.S. Department of Energy under
Contract No. DE-AC02-98CH10886.
The work of A.B. was supported by grants DOE DE-FC02-06ER-41439 and NSF 0555397.
A.V. work was supported by the Joint Theory Institute funded together by
Argonne National Laboratory and the University of Chicago, 
and in part by the U.S. Department of Energy, 
Division of High Energy Physics and Office of Nuclear Physics, under Contract DE-AC02-06CH11357.
\end{acknowledgments}

\end{document}